# LaIr$_3$Ga$_2$: A Superconductor based on a Kagome Lattice of Ir

*Xin Gui and Robert J. Cava\**

Department of Chemistry, Princeton University, Princeton, NJ 08540, USA

## *Abstract*

New materials based on Kagome lattices, predicted to host exotic quantum physics because they can display flat electronic bands, Dirac cones and topologically nontrivial surface states, are strongly desired. Here we report the crystal structure and superconducting properties of LaIr$_3$Ga$_2$, a previously unreported material that is based on a Kagome lattice of the heavy atom Ir. LaIr$_3$Ga$_2$ is a type II superconductor with $T_c$ ~ 5.2 K, $\mu_0 H_{c1}(0)$ ~ 7.1 mT and $\mu_0 H_{c2}(0)$ ~ 4.7 T. The normalized heat capacity jump at the superconducting transition, ($\Delta C/\gamma T_c$), 1.41, is within error of the value expected for a weak coupling BCS superconductor (1.43). Strong electron-electron correlation is inferred from the superconducting coherence length. Calculations show that the influence of spin orbit coupling on the electronic structure is significant, that the 5$d$ states of the Ir in the Kagome planes are dominant near the Fermi energy (E$_F$), that a nearly flat band is calculated to occur at about 100 meV below E$_F$, and that two Dirac points are close to E$_F$. This material is of interest for investigating the coupling between topological physics and superconductivity in a system with significant spin orbit coupling.

*Introduction*

Kagome lattice materials, made from corner-sharing triangles of magnetic atoms, have attracted much attention recent years. In addition to geometric magnetic frustration, the Kagome lattice structural motif can exhibit forefront quantum physics, such as the quantum spin liquid (QSL) state,[1–6] topological quantum state,[7–11] and superconductivity[12–19]. Flat bands, Dirac cones, and topologically nontrivial surface states are predicted and known for Kagome lattice materials[7,8,10,20–26], while insulating Kagome lattice compounds are well known frustrated magnets, such as $SrCr_{12-x}Ga_xO_{19}$ (SCGO)[27], and the QSL candidate $ZnCu_3(OH)Cl_2$.[28–30] In recent years, non-trivial electronic states in conducting Kagome-based systems have been explored in materials such as $Fe_3Sn_2$,[31] $Co_3Sn_2S_2$,[32] and $TbMn_6Sn_6$[33]. Directly related to this work, a novel family of layered materials based on a Kagome lattice of the light element vanadium, $AV_3Sb_5$ (A = K, Rb and Cs), has recently been found to host suprconductivity.[34] ($T_c$ = 0.93 K, 0.92 K and 2.5 K for the K, Rb and Cs variants, respectively).[14,35,36] One of the members of this family, $KV_3Sb_5$, was found to be a $\mathbb{Z}_2$ topological metal[14] and a giant anomalous Hall effect was observed at low temperature.[37] Both of these features were attributed to the presence of an unconventional chiral charge density wave.[12]

Another family of layered materials based on the Kagome lattice has also attracted attention. These intermetallic "132", $RT_3X_2$ phases, where R stands for a rare-earth element, T represents a transition metal and X is either Si or B[38,39] are derivatives of the hexagonal symmetry $CaCu_5$-type structure, with the $CaCu_2$ layer in that structure replaced by a $LaX_2$ layer. Some 132 type materials display a distorted transition metal Kagome lattice, while maintaining hexagonal symmetry.[13] Many superconductors have been found in this family, with the highest $T_c$ assigned to $LaRu_3Si_2$ ($T_c$ ~ 7.3 K).[13,15,38]. Nodeless Kagome superconductivity has been observed in that material by muon spin rotation (μSR) experiments and flat bands based on Ru $4d$ states were found a few meV above the Fermi energy.[13] These two families of superconductors, based on materials with $3d$ and $4d$ elements on their Kagome lattices, suggest that materials based on heavier, $5d$-element-based Kagome lattices, if found to be superconducting, would be strong candidates for investigating spin-orbit-coupling-dominated-physics on superconducting Kagome lattices.

Thus, here we report the novel superconductor $LaIr_3Ga_2$. This material crystallizes with the same hexagonal symmetry as $AV_3Sb_5$ and, unlike reported $LaRu_3Si_2$ which has a distorted Kagome lattice of Ru, possesses a perfect Kagome lattice made from the 5$d$ element Ir. The new material shows superconducting behavior below $T_c$ ~ 5.2 K, observed via by magnetization, resistivity and heat capacity measurements on arc-melted polycrystals. The superconducting parameters show a potentially stronger electron-electron interaction in LaIr$_3$Ga$_2$ than is reported for LaRu$_3$Si$_2$. Calculations are also presented to show that the electronic structure of the material is dominated by Ir 5$d$ states that exhibit the influence of spin orbit coupling.

## *Results and Discussion*

**Crystal Structure:** LaIr$_3$Ga$_2$ crystallizes in a CeCo$_3$B$_2$-type structure in the space group *P* 6/*mmm* (no. 191), which is the same as some other reported 132-phases, and the recently reported 135-superconductors.[34,38] The crystallographic data, including atomic positions, site occupancies, and refined anisotropic thermal displacement parameters are listed in Tables 1, 2 and 3. Single crystal X-ray diffraction patterns from the (hk0), (h0l) and (0kl) reciprocal lattice planes are shown in Figure S1 in the Supporting Information (SI); these support the high crystal quality. The crystal structure of LaIr$_3$Ga$_2$ is shown in Figure 1(a) and 1(b). The material adopts a simple layered structure where the two-dimensional Kagome lattice of Ir atoms is sandwiched by LaGa$_2$ layers. The layers are stacked along the crystal *c* axis. The In-plane Ir-Ir distance in the Kagome lattice plane is 2.7942 (2) Å, a little bit larger than the value observed in Ir metal (~2.714 Å). The La atoms form a triangular planar lattice with a separation of 5.5884 (4) Å. The Ga atoms, in the centers of the La triangles, construct a planar honeycomb lattice. Laboratory powder diffraction characterization of crushed as-melted polycrystalline material is shown in Figure S2. La$_2$Ir$_7$ and La$_5$Ir$_2$ are found as minor impurities. The lattice parameters of LaIr$_3$Ga$_2$ from Rietveld fitting of the powder diffraction data are *a* = 5.5830 (1) Å, *c* = 3.9108 (1) Å, consistent with the results from single crystal XRD. The refinement parameters $R_p$, $R_{wp}$ and $\chi^2$ are 5.68%, 7.78% and 2.46, respectively, indicating reliable refined results. A representative polycrystal sample is shown in Figure 1(c).

**Magnetic Properties:** The temperature-dependence of the magnetic susceptibility for LaIr$_3$Ga$_2$ was measured under multiple magnetic fields from 1.8 K to 7 K, as shown in the main panel of Figure 2(a). The diamagnetic state emerges below T$_c$ ~ 5.2 K. The diamagnetism is rapidly suppressed by low magnetic fields and almost disappears under the field of 100 mT, indicating a small lower critical field for this material. To most accurately characterize the magnetic properties of LaIr$_3$Ga$_2$, a demagnetization factor ($N$) was calculated by the equation $-b = \frac{1}{4\pi(1-N)}$ where b is the slope in M$_{fit}$ = bH and M$_{fit}$ is obtained by conducting linear fitting for the low-field region of the magnetization curve at 1.8 K in Figure 2(b). The resulting $N$ is 0.56 for LaIr$_3$Ga$_2$, consistent with the theoretical value for a cylindrical sample.[40] This $N$ value has been used to correct the magnetic susceptibility data in Figure 2(a). The inset of Figure 2(a) presents the corrected magnetic susceptibility of LaIr$_3$Ga$_2$ for both zero-field-cooling (ZFC) and field-cooling (FC) under an external magnetic field of 2 mT. The weak signal seen below the superconducting transition in the field-cooled data is exactly what is expected for flux pinning in a type II superconductor.

We have investigated the field-dependence of the magnetization (M$_v$) at various temperatures between 1.8 K and 5.4 K. With increasing magnetic field, M$_v$ rapidly decreases, with an upturn under a small magnetic field (~ 8 mT at 1.8 K, as shown in Figure 2(b)). To obtain a more accurate μ$_0$H$_{c1}$, M$_{fit}$, as mentioned above, is subtracted from M$_v$, as can be seen in Figure 2(c). A black dashed line M$_v$ – M$_{fit}$ = 0.2 emu/cm$^3$ is also shown in Figure 2(c), to indicate where the data points in Figure 2(d) have been extracted from. Note that the value 0.2 emu/cm$^3$ was determined to be less than 2% of M$_v$ at 1.8 K so that μ$_0$H$_{c1}$ can be estimated more accurately. The extracted data is fitted by using the equation below:

$$\mu_0 H_{c1}^*(T) = \mu_0 H_{c1}^*(0)\left[1 - \left(T/T_c\right)^2\right]$$

where $\mu_0 H_{c1}^*(0)$ is the lower critical field at 0 K and $T_c$ is superconducting transition temperature. The fitted lower critical field $\mu_0 H_{c1}^*(0)$ is 3.1 (1) mT and $T_{c\text{-}fit}$ = 5.3 (1) K, within experimental error of the temperature where the diamagnetism appears (Figure 2(a)). Moreover, when taking the demagnetization factor $N$ into account, the lower critical field can be obtained from:

$$\mu_0 H_{c1}(0) = \mu_0 H_{c1}^*(0)/(1-N)$$

Thus, the corrected lower critical field $\mu_0 H_{c1}(0)$ is 7.1 (2) mT.

**Resistivity:** In the inset of Figure 3(b), we show the temperature-dependence of the resistivity measured from 1.8 K to 300 K, for the same sample that was used for magnetic measurements. The sample shows metallic behavior with a residual-resistance ratio of ~3.4, which is either intrinsic or attributable to the sample's polycrystalline nature and grain boundaries. The low-temperature resistivity (7 K to 40 K) before entering superconducting state is fitted by using

$$\rho(T) = \rho_0 + AT^n$$

where $\rho_0$ is the residual resistivity due to defect scattering, $A$ is a constant and $n$ is an integer determined by the interaction pattern. Thus, for polycrystalline samples of LaIr$_3$Ga$_2$, $\rho_0$ is fitted to be 89.0 (4) μΩ cm, $A$ is 0.022 (1) μΩ cm/K$^3$ and $n$ is 2, which suggests that electron-electron interactions may dominate the resistivity between 7 K and 40 K. Figure 3(a) presents the temperature-dependence of resistivity measured from 1.8 K to 7 K under various magnetic fields. As is customary for such measurements, $T_c$ is defined as the temperature corresponding to the midpoint of the resistivity drop, as shown by the dashed arrow in Figure 3(a). In such measurements, T$_{c\text{-onset}}$ is ~5.95 K and zero resistivity is achieved at T$_{zero}$ ~ 5.28 K The $T_c$ is gradually suppressed by increasing magnetic field, and zero resistivity cannot be achieved at 1.8 K for μ$_0$H = 3 T and greater; for μ$_0$H = 4 T, no evidence for a superconducting transition can be seen in the resistivity above 1.8 K. The data points in Figure 3(b) are extracted from the intersections between dashed arrow and resistivity curves in Figure 3(a) and are utilized for fitting higher critical field using following Ginzburg-Landau relation:

$$\mu_0 H_{c2}(T) = \mu_0 H_{c2}(0) \frac{(1-t^2)}{(1+t^2)}$$

where t = T/$T_c$ and $T_c$ is a fitting parameter (transition temperature at zero magnetic field). The fitting line from GL relation well describes the observed superconducting behavior and results in $\mu_0 H_{c2}(0)$ = 4.71 (4) T and $T_{c\text{-fit}}$ = 5.16 (2) K. Based on BCS theory, the Pauli-limiting field for a superconductor can be described as $\mu_0 H_{c2}^p(0) = 1.85\,T_c$, that is, $\mu_0 H_{c2}^p(0)$ ~ 9.6 T for LaIr$_3$Ga$_2$, which is much larger than the experimental $\mu_0 H_{c2}(0)$.

The Ginzburg-Landau coherence length, $\xi_{GL}$ can be calculated from the given relation:

$$\mu_0 H_{c2}(0) = \frac{\Phi_0}{2\pi \xi_{GL}^2(0)}$$

where $\Phi_0$ = h/2e is the quantum flux and $\mu_0 H_{c2}(0)$ is as-measured-above. This makes the GL coherence length $\xi_{GL}(0)$ 84 Å, a zero-K GL coherence length that is very similar to that seen in heavy fermion superconductors, such as UPd$_2$Al$_3$ (85 Å),[41] and the previously reported 132-type superconductor LaRu$_3$Si$_2$ (107 Å).[19] This indicates that stronger electron-electron interaction exists in LaIr$_3$Ga$_2$ than is reported for LaRu$_3$Si$_2$ and that it is near that of heavy fermion superconductors. Moreover, the Ginzburg-Landau penetration depth at 0 K, $\lambda_{GL}(0)$, is calculated to be 2860 Å using given relation shown below with lower critical field $\mu_0 H_{c1}(0)$ determined previously

$$\mu_0 H_{c1}(0) = \frac{\Phi_0}{4\pi \lambda_{GL}^2(0)} \ln \frac{\lambda_{GL}(0)}{\xi_{GL}(0)}$$

The GL penetration depth found in LaIr$_3$Ga$_2$ is comparable to that previously reported to that for the noncentrosymmetric superconductor TaIr$_2$B$_2$ (3420 Å).[44] Furthermore, the Ginzburg-Landau parameter, $\kappa_{GL}(0) = \lambda_{GL}(0)/\xi_{GL}(0)$, can be calculated to be ~40, which is clearly larger than $1/\sqrt{2}$ implying that LaIr$_3$Ga$_2$ is a type-II superconductor. The thermodynamic critical field at 0 K ($\mu_0 H_c(0)$) is estimated as well by using the following relation

$$H_{c1}(0)H_{c2}(0) = H_c^2(0) \ln \kappa_{GL}(0)$$

yielding $\mu_0 H_c(0)$ = 97 mT for LaIr$_3$Ga$_2$.

**Heat Capacity:** After observation of both the Meissner effect in the magnetization and the zero resistivity, in order to fully confirm the bulk nature of the superconductivity in LaIr$_3$Ga$_2$, low-temperature heat capacity measurements were performed from 2 K to 10 K under applied magnetic fields of 0 T and 8 T, as shown in Figure 4(a). An obvious anomaly in the 0 applied field heat capacity, corresponding with the emergence of superconducting state, can be observed beginning at ~5.6 K, reaching a peak value at ~4.7 K, which is consistent with both magnetization and resistivity measurements. Under the magnetic field of 8 T, the superconducting heat capacity jump has disappeared consistent with the resistivity data, i.e., that the superconducting state is suppressed by the applied magnetic field in the measured temperature region. The normal-state (non-superconducting-state) heat capacity ($C_p$) - the heat capacity measured under 8 T in this case - often shows a linear relation between $C_p/T$ and $T^2$ according to the given relation

$$C_p/T = \gamma + \beta T^2$$

where $\gamma$ is the Sommerfeld coefficient, corresponding to electronic contribution to $C_p$, and $\beta$ is related to the phononic contribution. However, as shown in Figure S3 in the SI, this simple kind of relationship cannot describe the data from 2 K to 10 K, indicating that the simple Debye model cannot describe the material in this temperature range. Therefore, the following equation is employed to fit the normal-state heat capacity

$$C_p/T = \gamma + \beta T^2 + \eta T^4$$

where two terms, $\beta T^2$ and $\eta T^4$, are used to express the phononic contribution. Such fitting has been applied to some unconventional superconductors as well.[15] As shown by the green solid line in Figure 4(a), this relation matches the experimental data well and yields $\gamma = 15.0$ (2) mJ/mol/K$^2$, $\beta = 0.32$ (1) mJ/mol/K$^4$ and $\eta = 0.0082$ (1) mJ/mol/K$^6$. To estimate the heat capacity jump corresponding to the superconducting state, the normal state heat capacity measured under magnetic field of 8 T is subtracted from that measured under 0 T, as shown in Figure 4(b) By using the equal area construction, the normalized heat capacity jump ($\Delta C/\gamma T_c$) is calculated to be 1.41, close to the expected value of 1.43 for BCS superconductivity in the weak coupling limit.

Moreover, we calculate the Debye temperature using the given relation

$$\Theta_D = \left(\frac{12\pi^4}{5\beta}nR\right)^{1/3}$$

where $R$ is the gas constant with the value of 8.31 J/mol/K, $n$ is the number of atoms per formula unit ($n = 6$ for LaIr$_3$Ga$_2$). Thus, $\Theta_D$ is found to be 331 (5) K. With that information, the electron-phonon coupling constant $\lambda_{ep}$ can be estimated by using the inverted McMillan equation[43]

$$\lambda_{ep} = \frac{1.04 + \mu^* ln\left(\frac{\Theta_D}{1.45T_c}\right)}{(1 - 0.62\mu^*)ln\left(\frac{\Theta_D}{1.45T_c}\right) - 1.04}$$

where $\mu^*$ is the Coulomb pseudopotential parameter and is typically given a value of 0.13.[42,44–46] The value of $\lambda_{ep}$ is determined to be 0.63, indicating weakly coupled superconductivity. Finally, all the superconducting parameters obtained from experiments are summarized in Table 4.

**Electronic Structures:** To better interpret the properties of LaIr$_3$Ga$_2$, the electronic band structure, the orbital origin of its features, and the density of electronic states (DOS)) were calculated. The results of these calculations, with and without the inclusion of spin-orbit coupling (SOC) are plotted in Figure 5. By comparing Figure 5(a) & 5(b), various band splittings can be found throughout the Brillouin Zone (BZ) when SOC is included. When SOC is not involved, the calculated band structure of LaIr$_3$Ga$_2$ shows several intriguing features. Along the A-L direction, a Dirac point can be observed ~80 meV below the Fermi energy (E$_F$) while another Dirac point is seen exactly at E$_F$ along the H-A direction. Both of these Dirac points are gapped once SOC is taken into account. According to the projections of the La-$d$, Ir-$d$ and Ga-$p$ states on band structures, La-$d$ electrons are found near the A point around E$_F$ and along the Γ-M-K-Γ direction ~ 1.5 eV above E$_F$ while Ga-$p$ states are rarely found between -2 eV and 2 eV. However, the $d$ states from Ir atoms, i.e., the Kagome lattice, are dominant from -2 eV to ~ 0.15 eV, which suggests that the $d$ electrons of Ir play a significant role in determining the electronic properties of LaIr$_3$Ga$_2$. A nearly flat band dominated by Ir-*d* states can be observed after inclusion of SOC along the Γ-A direction at ~ -100 meV. A peak in the calculated electronic DOS may correspond to an instability, such as structural, electronic or magnetic instabilities. when SOC is involved, a broad peak presents at ~130 meV below E$_F$ which corresponds to the flat band along the Γ-A direction. Based on the DOS plots, SOC has a negligible effect on the DOS below ~ -0.5 eV and shows nearly no difference in the DOS (E$_F$), i.e., 4.87 states/eV/f.u. without SOC *vs* 4.84 states/eV/f.u. with SOC – thus the effect of spin orbit coupling is not simply to influence the total electronic DOS.

## *Conclusions*

In summary, we describe the structure and superconducting characteristics of a novel material, LaIr$_3$Ga$_2$, obtained by using the arc melting synthetic method. The new material has a layered structure in space group *P* 6/*mmm*, with an ideal highly symmetric Kagome lattice of Ir. Polycrystals were used to characterize the superconducting properties. Superconductivity was observed below a critical temperature of ~ 5.2 K, while lower and upper critical magnetic fields at 0 K were found to be ~7.1 mT and 4.7 T, respectively. The heat capacity measurements suggest weakly coupled BCS superconductivity with a relatively high Sommerfeld constant (about 15 mJ/mole/K$^2$) and the Ginzburg-Landau coherence length, $\xi_{GL}$, indicates stronger electron-electron correlation in LaIr$_3$Ga$_2$ than in LaRu$_3$Si$_2$. Electronic structure calculations of LaIr$_3$Ga$_2$ are consistent with the metallic behavior observed and reveal a nearly flat band ~100 meV below the Fermi energy when SOC effect is considered; two calculated Dirac points at E$_F$ are also observed while SOC is not included. This new material will be of interest for elucidating in the interactions between topologically nontrivial electronic states and superconductivity.

## *Experimental Details*

**Synthesis:** LaIr$_3$Ga$_2$ was synthesized by the arc-melting method. Elemental La (≥99.9%, ingot, Stream Chemicals), Ir (99.99%, ~22 mesh, Alfa Aesar) and Ga (99.99999%, ingot, Alfa Aesar) were utilized as starting materials. The La was handled in an Ar-filled glovebox. Iridium powder was first pressed into a pellet and arc melted to an Ir button. This step reduced the mass loss of Ir during the full synthesis. 100-150 mg of the starting materials with a molar ratio of 1:3.03:2.5 was then arc-melted twice under a high purity, Zr-gettered, argon atmosphere, with the button flipped over between melts. The resulting air-stable material was annealed at 500 °C for 2 days, but the phase purity and homogeneity did not change compared to that of the arc-melted buttons. Small polycrystalline samples with shiny faces were used in the property measurements.

**Crystal Structure:** Multiple small LaIr$_3$Ga$_2$ crystals (up to 80×80×10 μm$^3$) were tested by single crystal X-ray diffraction (XRD) to determine the crystal structure of the new material. The crystal structure, consistent among all crystals, was determined using a Bruker D8 QUEST single crystal diffractometer equipped with APEX III software and Mo radiation ($\lambda_{K\alpha}$= 0.71073 Å) at 300 K. The crystals were mounted on a Kapton loop protected by glycerol. Data acquisition was made *via* the Bruker SMART software with corrections for Lorentz and polarization effects included. A numerical absorption correction based on crystal-face-indexing was applied using *XPREP*. The direct method and full-matrix least-squares on F$^2$ procedure within the SHELXTL package were employed to solve the crystal structure.[47,48] Powder Xray diffraction patterns on crushed arc-melted buttons, obtained with a Bruker D8 Advance Eco diffractometer with Cu Kα radiation and a LynxEye-XE detector, were consistent with the structure determined by single crystal diffraction.

**Physical Property Measurement:** The DC magnetization was measured from 1.8 to 300 K under various applied magnetic fields using a Quantum Design Dynacool Physical Property Measurement System (PPMS), equipped with a vibrating sample magnetometer (VSM) option. The magnetic susceptibility was defined as M(emu)/H(Oe). Field-dependent magnetization data was collected at different temperatures with applied magnetic fields ranging from $\mu_0 H$ = -9 T to 9 T. The resistivity measurements were carried out in in the same system using the four-probe method between 1.8 K and 300 K under magnetic fields up to $\mu_0 H$ = 9 T. Platinum wires were attached to the samples by silver epoxy to ensure ohmic contact. Heat capacity was measured using a standard relaxation method in the PPMS.

**Electronic Structure Calculations:** The electronic structure and electronic density of states (DOS) of LaIr$_3$Ga$_2$ were calculated using WIEN2k, which employs the full-potential linearized augmented plane wave method (FP-LAPW) with local orbitals implemented.[49,50] The electron exchange-correlation potential used was the generalized gradient approximation.[51] The conjugate gradient algorithm was applied. Reciprocal space integrations were completed over a 5×5×6 Monkhorst-Pack *k*-point mesh.[52] Spin-orbit coupling (SOC) was applied to both the La and Ir atoms. The lattice parameters and atomic positions obtained from single crystal XRD were used in all calculations. With these settings, the calculated total energy converged to less than 0.1 meV per atom.

## *Data Availability*

The data that support the findings of this study are available from the corresponding authors upon reasonable request.

## *References*

1. Yan, S., Huse, D. A. & White, S. R. Spin-Liquid Ground State of the S = 1/2 Kagome Heisenberg Antiferromagnet. *Science* **332**, 1173–1176 (2011).
2. Han, T.-H. *et al.* Fractionalized excitations in the spin-liquid state of a kagome-lattice antiferromagnet. *Nature* **492**, 406–410 (2012).
3. Fu, M., Imai, T., Han, T.-H. & Lee, Y. S. Evidence for a gapped spin-liquid ground state in a kagome Heisenberg antiferromagnet. *Science* **350**, 655–658 (2015).
4. Lee, S.-H. *et al.* Quantum-spin-liquid states in the two-dimensional kagome antiferromagnets $Zn_xCu_{4-x}(OD)_6Cl_2$. *Nat. Mater.* **6**, 853–857 (2007).
5. Zhou, Y., Kanoda, K. & Ng, T.-K. Quantum spin liquid states. *Rev. Mod. Phys.* **89**, 025003 (2017).
6. Chamorro, J. R., McQueen, T. M. & Tran, T. T. Chemistry of Quantum Spin Liquids. *Chem. Rev.* **121**, 2898–2934 (2021).
7. Xue, H., Yang, Y., Gao, F., Chong, Y. & Zhang, B. Acoustic higher-order topological insulator on a kagome lattice. *Nat. Mater.* **18**, 108–112 (2019).
8. Chisnell, R. *et al.* Topological Magnon Bands in a Kagome Lattice Ferromagnet. *Phys. Rev. Lett.* **115**, 147201 (2015).
9. Guguchia, Z. *et al.* Tunable anomalous Hall conductivity through volume-wise magnetic competition in a topological kagome magnet. *Nat. Commun.* **11**, 559 (2020).
10. Zhang, X., Jin, L., Dai, X. & Liu, G. Topological Type-II Nodal Line Semimetal and Dirac Semimetal State in Stable Kagome Compound $Mg_3Bi_2$. *J. Phys. Chem. Lett.* **8**, 4814–4819 (2017).
11. Zhou, H. *et al.* Enhanced anomalous Hall effect in the magnetic topological semimetal $Co_3Sn_{2-x}In_xS_2$. *Phys. Rev. B* **101**, 125121 (2020).
12. Jiang, Y.-X. *et al.* Unconventional chiral charge order in kagome superconductor $KV_3Sb_5$. *Nat. Mater.* **20**, 1353–1357 (2021).

13. Mielke, C. *et al.* Nodeless kagome superconductivity in LaRu$_3$Si$_2$. *Phys. Rev. Mater.* **5**, 034803 (2021).

14. Ortiz, B. R. *et al.* Superconductivity in the Z$_2$ kagome metal KV$_3$Sb$_5$. *Phys. Rev. Mater.* **5**, 034801 (2021).

15. Li, S. *et al.* Anomalous properties in the normal and superconducting states of LaRu$_3$Si$_2$. *Phys. Rev. B* **84**, 214527 (2011).

16. Tan, H., Liu, Y., Wang, Z. & Yan, B. Charge density waves and electronic properties of superconducting kagome metals. *Phys. Rev. Lett.* **127**, 046401 (2021).

17. Wang, Z. *et al.* Electronic nature of chiral charge order in the kagome superconductor CsV$_3$Sb$_5$. *Phys. Rev. B* **104**, 075148 (2021).

18. Chen, H. *et al.* Roton pair density wave in a strong-coupling kagome superconductor. *Nature* 1–9 (2021) doi:10.1038/s41586-021-03983-5.

19. Kishimoto, Y. *et al.* Magnetic Susceptibility Study of LaRu$_3$Si$_2$. *J. Phys. Soc. Jpn.* **71**, 2035–2038 (2002).

20. Yin, J.-X. *et al.* Negative flat band magnetism in a spin–orbit-coupled correlated kagome magnet. *Nat. Phys.* **15**, 443–448 (2019).

21. Kang, M. *et al.* Topological flat bands in frustrated kagome lattice CoSn. *Nat. Commun.* **11**, 4004 (2020).

22. Li, M. *et al.* Dirac cone, flat band and saddle point in kagome magnet YMn$_6$Sn$_6$. *Nat. Commun.* **12**, 3129 (2021).

23. Han, M. *et al.* Evidence of two-dimensional flat band at the surface of antiferromagnetic kagome metal FeSn. *Nat. Commun.* **12**, 5345 (2021).

24. Kang, M. *et al.* Dirac fermions and flat bands in the ideal kagome metal FeSn. *Nat. Mater.* **19**, 163–169 (2020).

25. Ma, D.-S. *et al.* Spin-Orbit-Induced Topological Flat Bands in Line and Split Graphs of Bipartite Lattices. *Phys. Rev. Lett.* **125**, 266403 (2020).

26. Peri, V., Song, Z.-D., Bernevig, B. A. & Huber, S. D. Fragile Topology and Flat-Band Superconductivity in the Strong-Coupling Regime. *Phys. Rev. Lett.* **126**, 027002 (2021).

27. Aeppli, G., Broholm, C., Ramirez, A., Espinosa, G. P. & Cooper, A. S. Broken spin rotation symmetry without magnetic Bragg peaks in Kagomé antiferromagnets. *J. Magn. Magn. Mater.* **90–91**, 255–259 (1990).

28. Mendels, P. & Bert, F. Quantum Kagome Antiferromagnet $ZnCu_3(OH)_6Cl_2$. *J. Phys. Soc. Jpn.* **79**, 011001 (2010).

29. Shores, M. P., Nytko, E. A., Bartlett, B. M. & Nocera, D. G. A Structurally Perfect S = 1/2 Kagomé Antiferromagnet. *J. Am. Chem. Soc.* **127**, 13462–13463 (2005).

30. Helton, J. S. *et al.* Spin Dynamics of the Spin-$1/2$ Kagome Lattice Antiferromagnet $ZnCu_3(OH)_6Cl_2$. *Phys. Rev. Lett.* **98**, 107204 (2007).

31. Ye, L. *et al.* Massive Dirac fermions in a ferromagnetic kagome metal. *Nature* **555**, 638–642 (2018).

32. Morali, N. *et al.* Fermi-arc diversity on surface terminations of the magnetic Weyl semimetal $Co_3Sn_2S_2$. *Science* **365**, 1286–1291 (2019).

33. Yin, J.-X. *et al.* Quantum-limit Chern topological magnetism in $TbMn_6Sn_6$. *Nature* **583**, 533–536 (2020).

34. Ortiz, B. R. *et al.* New kagome prototype materials: discovery of $KV_3Sb_5$, $RbV_3Sb_5$, and $CsV_3Sb_5$. *Phys. Rev. Mater.* **3**, 094407 (2019).

35. Yin, Q. *et al.* Superconductivity and Normal-State Properties of Kagome Metal $RbV_3Sb_5$ Single Crystals. *Chin. Phys. Lett.* **38**, 037403 (2021).

36. Ortiz, B. R. *et al.* $CsV_3Sb_5$: A $Z_2$ Topological Kagome Metal with a Superconducting Ground State. *Phys. Rev. Lett.* **125**, 247002 (2020).

37. Yang, S.-Y. *et al.* Giant, unconventional anomalous Hall effect in the metallic frustrated magnet candidate, $KV_3Sb_5$. *Sci. Adv.* **6**, eabb6003.

38. Ku, H. C., Meisner, G. P., Acker, F. & Johnston, D. C. Superconducting and magnetic properties of new ternary borides with the $CeCo_3B_2$-type structure. *Solid State Commun.* **35**, 91–96 (1980).

39. Athreya, K. S. *et al.* Superconductivity in the ternary borides $CeOs_3B_2$ and $CeRu_3B_2$: Magnetic susceptibility and specific heat measurements. *Phys. Lett. A* **113**, 330–334 (1985).

40. Sato, M. & Ishii, Y. Simple and approximate expressions of demagnetizing factors of uniformly magnetized rectangular rod and cylinder. *J. Appl. Phys.* **66**, 983–985 (1989).

41. Geibel, C. *et al.* Heavy-fermion superconductivity at $T_c=2K$ in the antiferromagnet $UPd_2Al_3$. *Z. Phys. B - Condens. Matter* **84**, 1–2 (1991).

42. Górnicka, K. *et al.* $NbIr_2B_2$ and $TaIr_2B_2$ – New Low Symmetry Noncentrosymmetric Superconductors with Strong Spin–Orbit Coupling. *Adv. Funct. Mater.* **31**, 2007960 (2021).

43. McMillan, W. L. Transition Temperature of Strong-Coupled Superconductors. *Phys. Rev.* **167**, 331–344 (1968).

44. Carnicom, E. M. *et al.* $TaRh_2B_2$ and $NbRh_2B_2$: Superconductors with a chiral noncentrosymmetric crystal structure. *Sci. Adv.* **4**, eaar7969.

45. Verchenko, V. Yu., Tsirlin, A. A., Zubtsovskiy, A. O. & Shevelkov, A. V. Strong electron-phonon coupling in the intermetallic superconductor $Mo_8Ga_{41}$. *Phys. Rev. B* **93**, 064501 (2016).

46. Singh, D., Hillier, A. D., Thamizhavel, A. & Singh, R. P. Superconducting properties of the noncentrosymmetric superconductor $Re_6Hf$. *Phys. Rev. B* **94**, 054515 (2016).

47. Sheldrick, G. M. SHELXT – Integrated space-group and crystal-structure determination. *Acta Cryst. A* **71**, 3–8 (2015).

48. Walker, N. & Stuart, D. An empirical method for correcting diffractometer data for absorption effects. *Acta Cryst. A* **39**, 158–166 (1983).

49. Wimmer, E., Krakauer, H., Weinert, M. & Freeman, A. J. Full-potential self-consistent linearized-augmented-plane-wave method for calculating the electronic structure of molecules and surfaces: $O_2$ molecule. *Phys. Rev. B* **24**, 864–875 (1981).

50. Schwarz, K. & Blaha, P. Solid state calculations using WIEN2k. *Comput. Mater. Sci.* **28**, 259–273 (2003).

51. Perdew, J. P. & Wang, Y. Accurate and simple analytic representation of the electron-gas correlation energy. *Phys. Rev. B* **45**, 13244–13249 (1992).

52. King-Smith, R. D. & Vanderbilt, D. Theory of polarization of crystalline solids. *Phys. Rev. B* **47**, 1651–1654 (1993).

*Acknowledgements*

This research was supported by the Gordon and Betty Moore Foundation, EPIQS initiative, grant GBMF-9066. The authors acknowledge helpful conversations with Evgueni Talantsev at the M.N. Miheev Institute of Metal Physics in Ekaterinburg Russia, about the interpretation of the data.

## *Author Information*

### *Affiliations*

Department of Chemistry, Princeton University, Princeton, NJ 08540, USA

Xin Gui & Robert J. Cava

### *Contributions*

X. G. and R. J. C. designed the project. X. G. performed experiments and theoretical calculations. X. G. and R. J. C. discussed the results and wrote the manuscript.

### *Corresponding Author*

Correspondence to Robert J. Cava: rcava@princeton.edu

## *Ethics declarations*

The authors declare no competing interests.

**Table 1.** Single crystal structure refinement for LaIr$_3$Ga$_2$ at 301 (2) K.

| Refined Formula | LaIr$_3$Ga$_2$ |
|---|---|
| F.W. (g/mol) | 854.95 |
| Space group; Z | P 6/*mmm*; 1 |
| $a$(Å) | 5.5884 (3) |
| $c$(Å) | 3.9063 (3) |
| V (Å$^3$) | 105.65 (1) |
| Extinction Coefficient | 0.010 (1) |
| θ range (º) | 4.211-30.979 |
| No. reflections; $R_{int}$ | 1393; 0.0217 |
| No. independent reflections | 91 |
| No. parameters | 9 |
| $R_1$: $\omega R_2$ ($I>2\delta(I)$) | 0.0124; 0.0339 |
| Goodness of fit | 1.182 |
| Diffraction peak and hole (e$^-$/ Å$^3$) | 1.356; -1.836 |

**Table 2.** Atomic coordinates and equivalent isotropic displacement parameters for LaIr$_3$Ga$_2$ at 301 (2) K. (U$_{eq}$ is defined as one-third of the trace of the orthogonalized U$_{ij}$ tensor (Å$^2$))

| Atom | Wyck. | Occ. | $x$ | $y$ | $z$ | $U_{eq}$ |
|---|---|---|---|---|---|---|
| Ir1 | 3$g$ | 1 | ½ | 0 | ½ | 0.0083 (2) |
| La2 | 1$a$ | 1 | 0 | 0 | 0 | 0.0103 (2) |
| Ga3 | 2$c$ | 1 | ⅓ | ⅔ | 0 | 0.0057 (3) |

**Table 3.** Anisotropic thermal displacement parameters for LaIr$_3$Ga$_2$.

| Atom | U11 | U22 | U33 | U12* | U13* | U23* |
|---|---|---|---|---|---|---|
| Ir1 | 0.0100 (2) | 0.0046 (2) | 0.0083 (3) | 0.0023 (1) | 0 | 0 |
| La2 | 0.0072 (3) | 0.0072 (3) | 0.0167 (5) | 0.0036 (2) | 0 | 0 |
| Ga3 | 0.0060 (3) | 0.0060 (3) | 0.0050 (5) | 0.0030 (2) | 0 | 0 |

**Table 4.** Observed superconducting parameters of LaIr$_3$Ga$_2$.

| Parameter | Unit | Value |
|---|---|---|
| $T_c$ | K | 5.16 |
| $\mu_0 H_{c1}(0)$ | mT | 7.1 |
| $\mu_0 H_{c2}(0)$ | T | 4.71 |
| $\mu_0 H_c(0)$ | mT | 97 |
| $\mu_0 H_{c2}^p(0)$ | T | 9.6 |
| $\xi_{GL}(0)$ | Å | 84 |
| $\lambda_{GL}(0)$ | Å | 2860 |
| $\kappa_{GL}(0)$ | \ | 34 |
| $\gamma$ | mJ/mol/K$^2$ | 15.0 |
| $\Delta C/\gamma T_c$ | \ | 1.41 |
| $\Theta_D$ | K | 331 |
| $\lambda_{ep}$ | \ | 0.63 |

**Figure 1.** Crystal structure of LaIr$_3$Ga$_2$ viewed from **(a)** (110) direction and **(b)** $c$ axis. Red, blue and orange balls represent La, Ir and Ga atoms, respectively. (c) Picture of sample with shiny faces used for physical properties measurements.

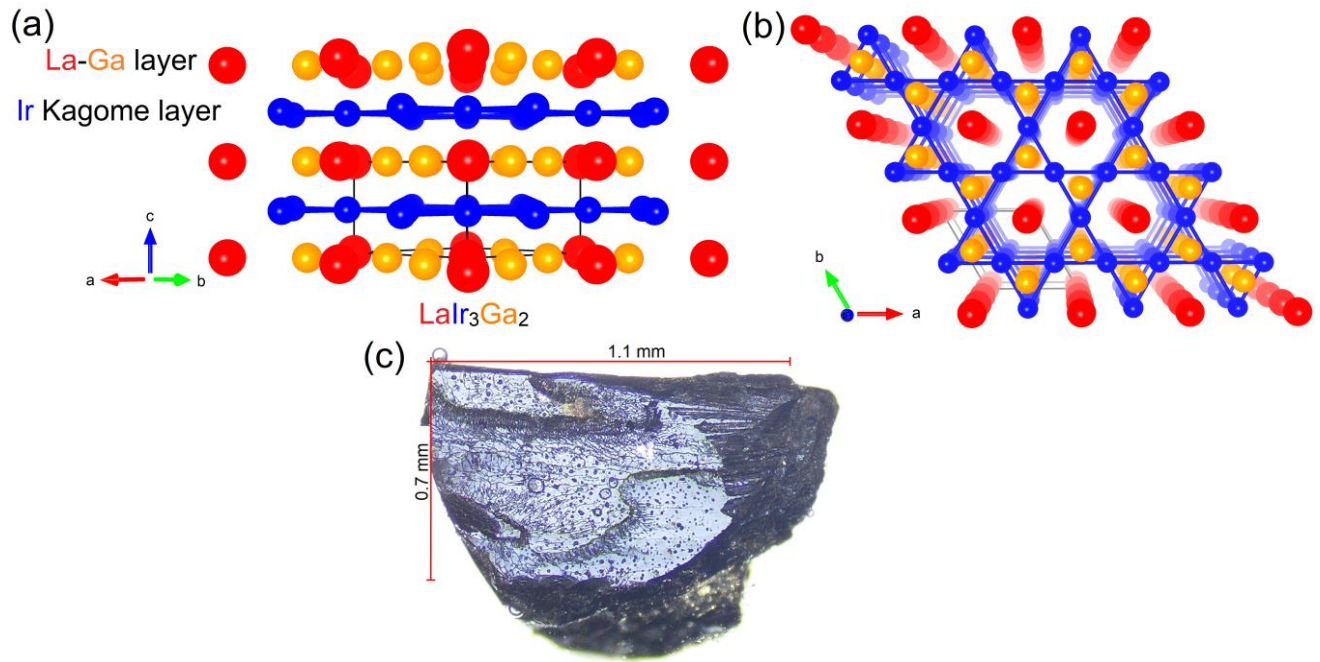

**Figure 2. (a) (Main panel)** Temperature-dependence of volume magnetic susceptibility corrected for demagnetization effect for LaIr$_3$Ga$_2$ from 1.8 K to 7 K under multiple magnetic fields. **(Inset)** Temperature-dependence of volume magnetic susceptibility from 2 K to 10 K under magnetic field of 2 mT through both zero-field-cooling (ZFC) and field-cooling (FC). **(b)** Field-dependence of volume magnetization ($M_v$) from 0 to 30 mT under various temperatures from 1.8 K to 5.4 K. The fitting for demagnetization factor ($M_{fit}$) is shown as black solid line. **(c)** Difference between $M_v$ and $M_{fit}$ at lower field (0-10 mT) under multiple temperatures. **(d)** Lower critical field ($\mu_0H_{c1}$) fitting while the data points are taken from the intersections between dashed black line and colored curves in **(c)**.

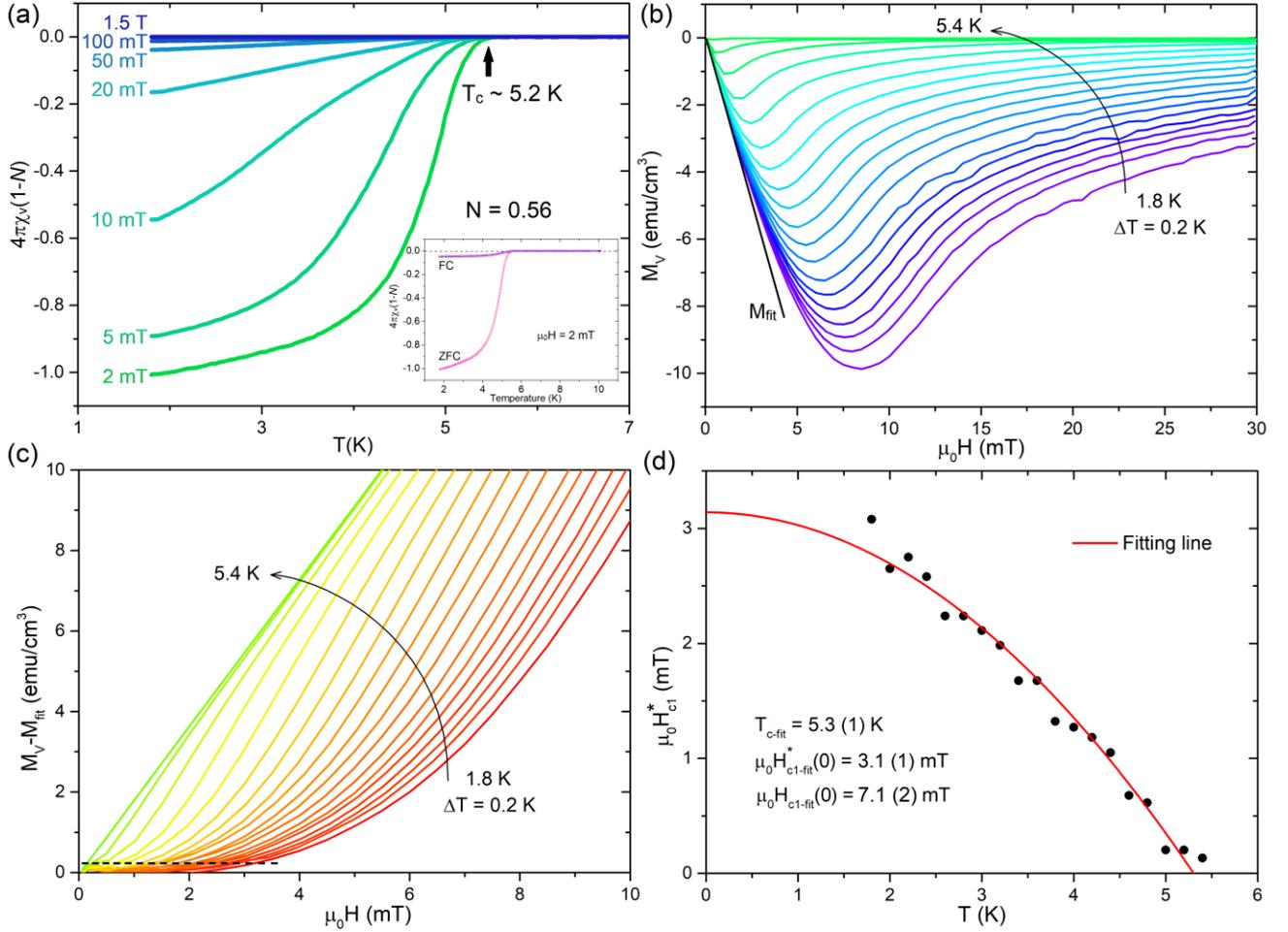

**Figure 3. (a)** Temperature-dependence of the resistivity from 1.8 K to 7 K under different magnetic fields from 0 T to 4 T. **(b) (Main panel)** Temperature-dependence of upper critical field $\mu_0H_{c2}$ (black solid circles) and the upper critical field fitting (red solid line). **(Inset)** Temperature-dependence of resistivity from 1.8 K to 300 K with fitting (cyan solid line).

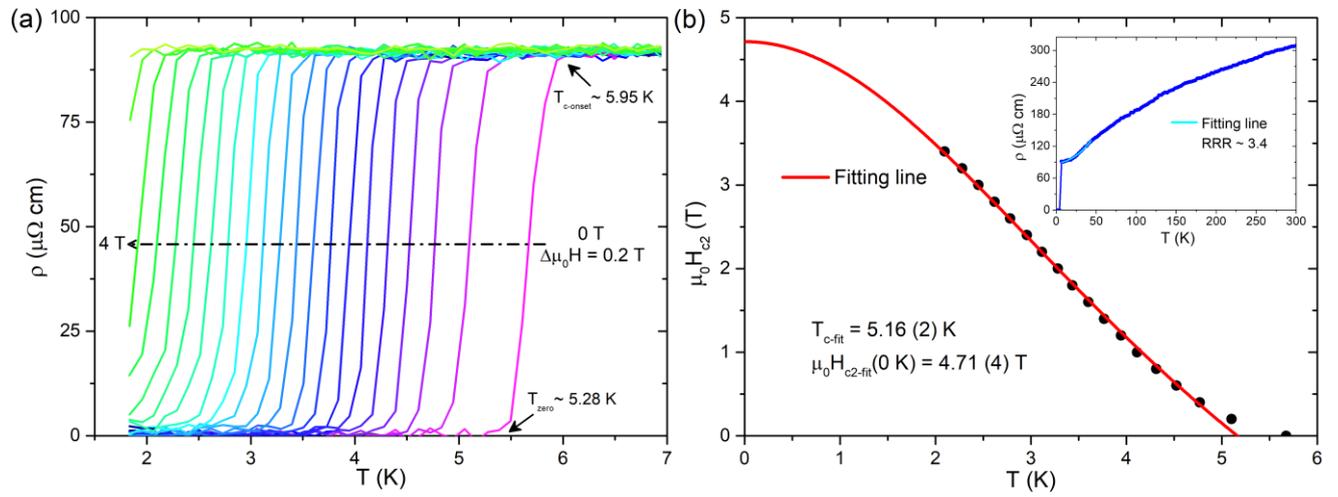

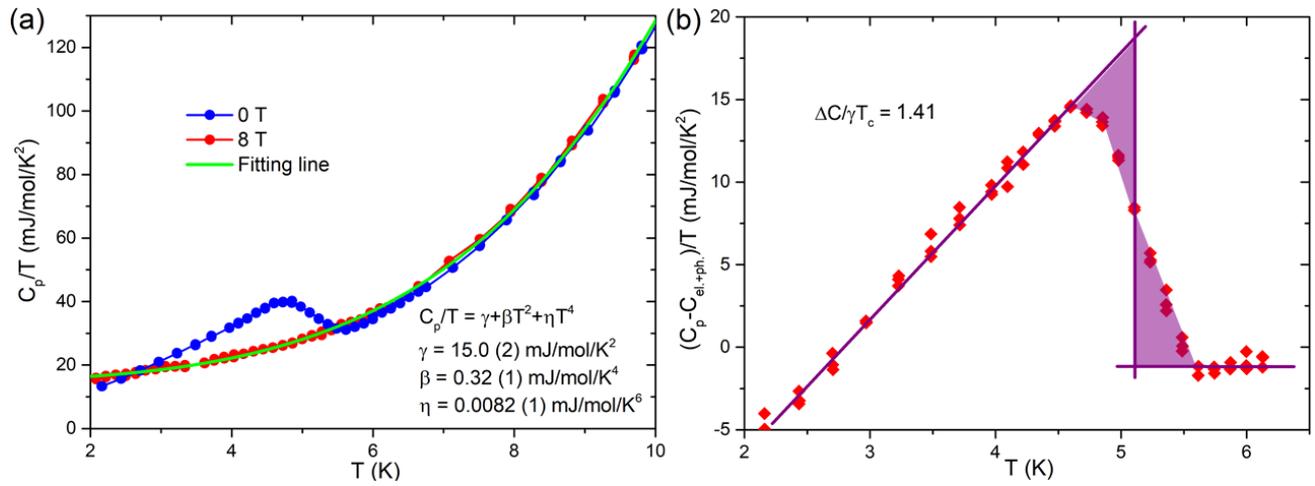

**Figure 4. (a)** $C_p/T$ vs $T$ measured under magnetic fields of 0 T and 8 T. Heat capacity jump corresponding to superconductivity emerges under 0 T while it is suppressed under 8 T. Green solid line stands for the fitting for both electronic ($C_{el.}$) and phononic ($C_{ph.}$) contribution under normal state. **(b)** $(C_p-C_{el.+ph.})/T$ vs $T$ from 2 K to 6 K. The purple solid lines and shadows are guide for the eye for the equal area construction used for estimation of $T_c$ and the superconducting jump $\Delta C/T_c$.

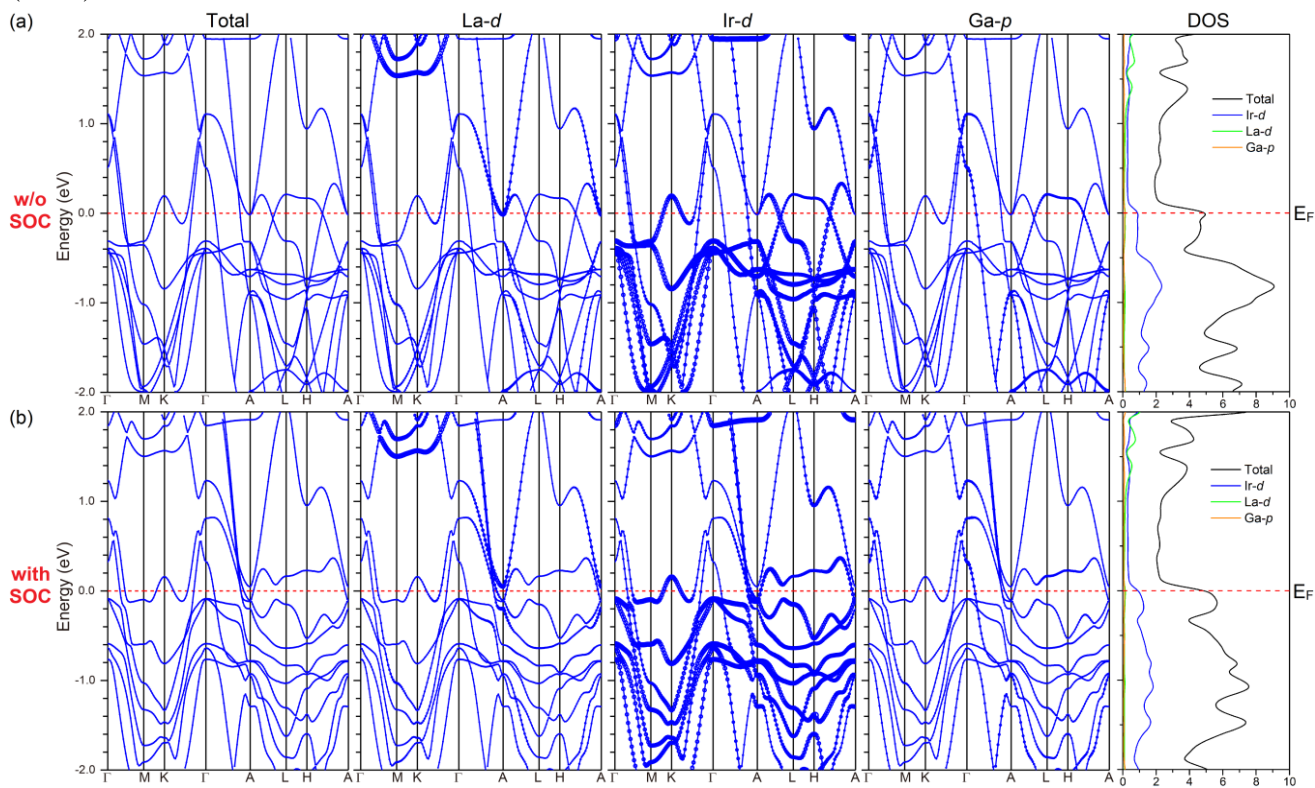

**Figure 5.** The calculated electronic structure in the vicinity (from -2 eV to 2 eV) of the Fermi Energy (E$_F$), orbital projections of band structure (*p* states of Ga and *d* states of La and Ir) and the calculated electronic density of states of LaIr$_3$Ga$_2$ **(a)** without and **(b)** with the inclusion of spin-orbit coupling (SOC). The red dashed line indicates E$_F$.

**Supporting Information for**

# LaIr$_3$Ga$_2$: A Superconductor based on a Kagome Lattice of Ir

*Xin Gui and Robert J. Cava\**

Department of Chemistry, Princeton University, Princeton, NJ 08540, USA

**Table of Contents**

**Figure S1.** Single crystal X-ray diffraction patterns for the (0kl), (h0l) and (hk0) reciprocal planes of LaIr$_3$Ga$_2$.

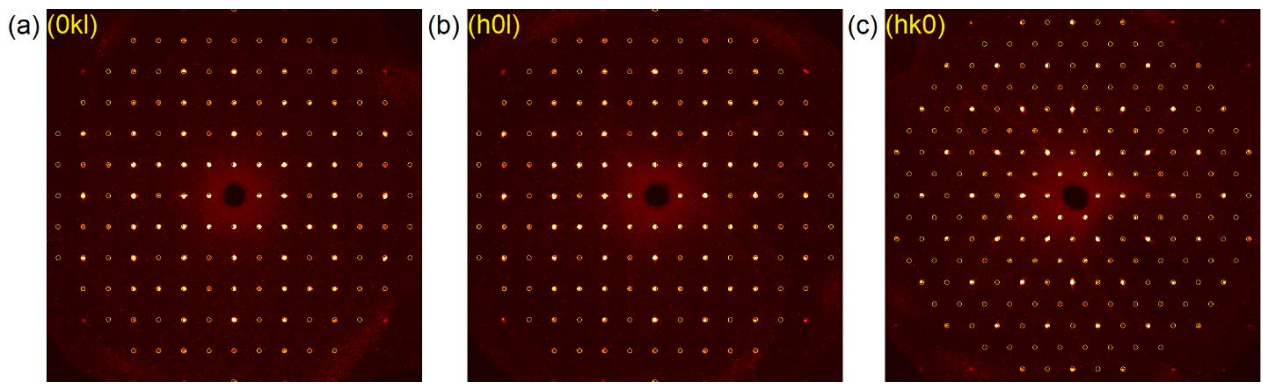

**Figure S2.** Powder X-ray diffraction pattern of a bulk sample of nominal composition LaIr$_3$Ga$_2$ with Rietveld fitting to the crystal structure. The black points are the observed pattern. The red and blue lines indicate calculated and difference between observed and calculated patterns, respectively. Vertical green, orange and pink sticks represent Bragg peak positions for LaIr$_3$Ga$_2$, La$_2$Ir$_7$ and La$_5$Ir$_2$.

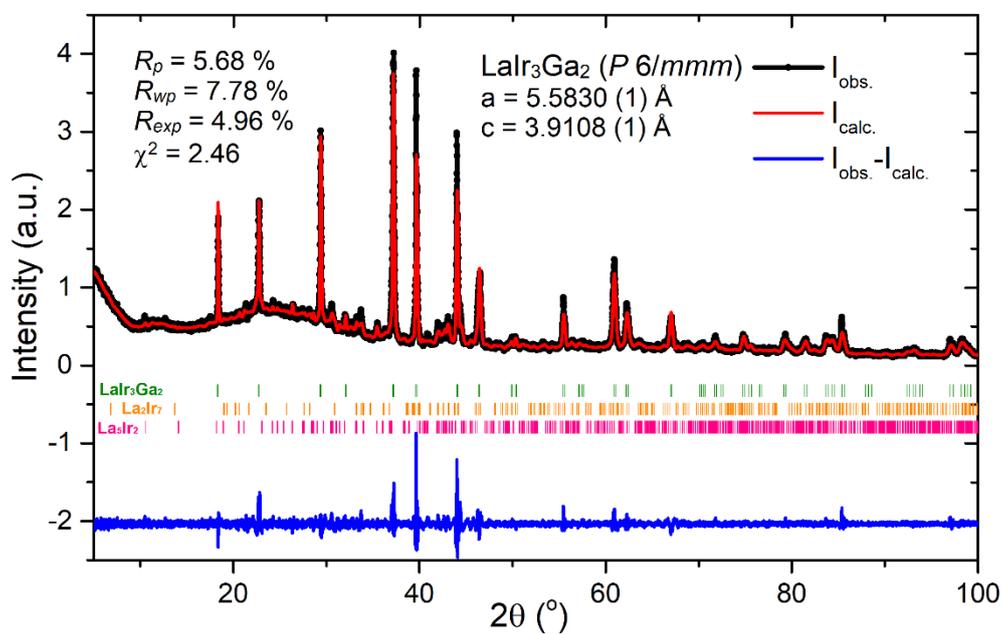

**Figure S3.** $C_p/T$ vs $T^2$ measured under magnetic field of 8 T from 2 K to 10 K. A linear relation based on the Debye model $C_p/T = \gamma + \beta T^2$ cannot be found in this temperature range.

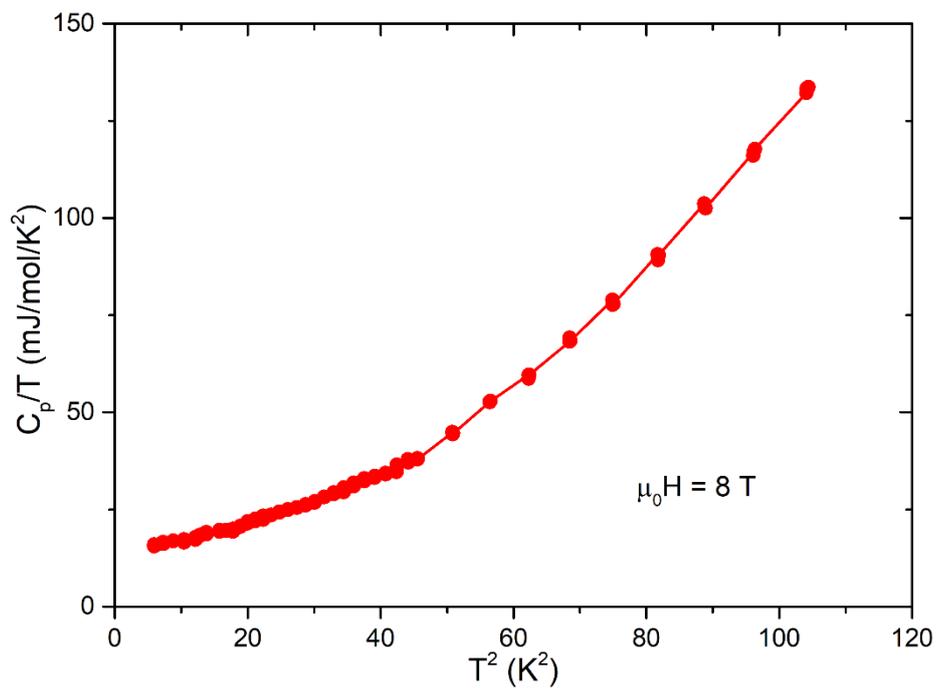